\documentclass[aps,prl,floatfix,showpacs,twocolumn]{revtex4-1}
\usepackage{ulem}
\usepackage{dcolumn}
\usepackage{bm}
\usepackage{color}
\usepackage{amssymb}
\usepackage{amsmath}
\usepackage{graphicx}
\usepackage{amsfonts}
\usepackage{slashed}
\usepackage{pstricks}
\usepackage{float}
\usepackage{hyperref}
\usepackage{array}
\allowdisplaybreaks
\usepackage{dcolumn}
\usepackage{epsf}
\usepackage{tabularx}
\usepackage{soul}
\usepackage{siunitx}
\usepackage{wrapfig}

\begin{document}
\title{Muon capture in nuclei: an {\itshape ab initio} approach based on quantum Monte Carlo methods}
\author{
A.\ Lovato$^{\, {\rm a,b} }$,
N.\ Rocco$^{\, {\rm b,c} }$, and
R.\ Schiavilla$^{\, {\rm d,e} }$
}
\affiliation{
$^{\,{\rm a}}$\mbox{INFN-TIFPA Trento Institute of Fundamental Physics and Applications, 38123 Trento, Italy}\\
$^{\,{\rm b}}$\mbox{Physics Division, Argonne National Laboratory, Argonne, IL 60439, USA}\\
$^{\,{\rm c}}$\mbox{Theoretical Physics Department, Fermi National Accelerator Laboratory, Batavia, IL 60510, USA}\\
$^{\,{\rm d}}$\mbox{Department of Physics, Old Dominion University, Norfolk, VA 23529, USA}\\
$^{\,{\rm e}}$\mbox{Theory Center, Jefferson Lab, Newport News, VA 23606, USA}
}
\date{\today}

%
\date{\today}
\begin{abstract}
{An {\it ab initio} quantum Monte Carlo method is introduced for calculating total rates of muon weak
capture in light nuclei with mass number $A \leq 12$.  As a first application of the method, we perform
a calculation of the rate in $^4$He in a dynamical framework based on realistic two- and three-nucleon
interactions and realistic nuclear charge-changing weak currents.  The currents include one- and two-body terms
induced by $\pi$- and $\rho$-meson exchange, and $N$-to-$\Delta$ excitation, and are constrained
to reproduce the empirical value of the Gamow-Teller matrix element in tritium.  We investigate
the sensitivity of theoretical predictions to current parametrizations of the nucleon axial and induced
pseudoscalar form factors as well as to two-body contributions in the weak currents.  The large
uncertainties in the measured values obtained from bubble-chamber experiments
(carried out over 50 years ago) prevent us from drawing any definite conclusions.
}
\end{abstract}
\maketitle
Negative muons passing through matter can be captured into high-lying atomic orbitals,
from where they rapidly cascade down into the $1s$ orbital.  There, they either decay via the process
$\mu^- \rightarrow e^- \,\overline{\nu}_e\, \nu_\mu$ with a rate which is almost
the same as in free space~\cite{Czarnecki:2000}, or are captured by the nucleus in a
weak-interaction process resulting in the change of one of the protons into a neutron
at a rate that is proportional to $Z^4$~\cite{Primakoff:1959}, where $Z$ is the nucleus'
proton number, and which, at least for light nuclei, is much smaller than the free decay
rate. 

In the nuclear capture, the muon rest mass ($m_\mu$) is converted in energy shared by the
emitted (muon) neutrino and recoiling final nucleus.  Since $m_\mu\,$$\,\approx$$\, 105$ MeV,
a calculation of the total inclusive rate---i.e., summed over all final states---requires, in principle,
knowledge of both the low-lying discrete states and higher-energy continuum spectrum of the
final nucleus.  In {\it ab initio} dynamical approaches based on realistic nuclear interactions, the
solution of the scattering problem poses a significant challenge, even for capture in nuclei as light
as $^3$He and $^3$H.  Indeed, while accurate theoretical estimates of the $^3$He$(\mu^-,\nu_\mu)^3$H
rate (a transition only involving bound states) have been made since the early 1990's~\cite{Congleton:1992,
Congleton:1995,Marcucci:2002,Marcucci:2012}, it is only recently that studies based on the Faddeev
method and accounting for the contributions to the rate from the breakup channels of $^3$He
(into $^2$H+n and $^1$H+$2\,n$) and $^3$H (into $3\,n$) have appeared in the literature, respectively in
Refs.~\cite{Golak:2014} and~\cite{Golak:2016}.  

The other important aspect of muon capture has to do with the description of the nuclear charge-changing
weak current responsible for the $p$-$n$ conversion. Its dominant one-body term is associated
with the matrix element $\langle n|\overline{d}\, \gamma^\mu(1-\gamma_5)|p\rangle$, and is
parametrized in terms of four form factors (FFs).  Two of these, $F_1(q^2)$ and $F_2(q^2)$ ($q^2$
is the lepton four-momentum transfer), enter the vector component, and are related to the
isovector electromagnetic FFs by the conserved-vector-current (CVC) constraint.  The remaining two,
the axial and induced pseudoscalar FFs, respectively $G_A(q^2)$ and $G_{PS}(q^2)$, characterize
the axial component.  The $F_1(q^2)$ and $F_2(q^2)$ FFs are well known over a broad range of momentum
transfers from elastic electron scattering off protons and deuterons~\cite{Hyde:2004}.  The value $g_A$ of
the axial FF at vanishing $q^2$ is precisely determined from neutron $\beta$ decay,
$g_A\,$=$\,1.2723 (23)$~\cite{Patrignani:2016}, while the $q^2$-dependence is parametrized by
a dipole form with a cutoff $\Lambda_A\,$$\approx$$\, 1$ GeV as obtained in analyses of pion
electroproduction data~\cite{Amaldi:1979} and direct measurements of
$\nu_\mu/\overline{\nu}_\mu$-$p$~\cite{Ahrens:1987} and quasielastic
$\nu_\mu$-$d$~\cite{Baker:1981,Miller:1982,Kitagaki:1983} scattering cross sections.
A recent measurement of muon capture in hydrogen by the MuCap
collaboration at PSI~\cite{Andreev:2013} has led to a precise determination of the
$G_{PS}(q^2)$ FF (the least well experimentally known of the four), $G_{PS}(-0.88\, m_\mu^2)=8.06\pm 0.55$,
a value that is consistent with theoretical predictions derived from
chiral perturbation theory~\cite{Bernard:1994,Bernard:2002}.

In the nuclear charge-changing weak current, in addition to one-body, there are two-body terms that
arise quite naturally in the conventional meson-exchange picture, for reviews
see Refs.~\cite{Towner:1987,Riska:1988}, as
well as in more modern approaches based on chiral effective field
theory~\cite{Park:1993,Park:1996,Pastore:2009,Pastore:2011,Koelling:2009,Piarulli:2013,Baroni:2016,Krebs:2017}.
Those in the vector sector are related by CVC to the isovector two-body
electromagnetic currents, notably the long-range currents induced by pion exchange.  By now, there is a
substantial body of experimental evidence for their presence from a variety of photo- and electro-nuclear
transitions in nuclei, including, among others, thermal neutron radiative captures on hydrogen and helium isotopes,
magnetic moments and $M1$ transition rates in light nuclei, elastic and transition magnetic
form factors of few-nucleon systems, and lastly transverse response functions measured in
quasielastic $(e,e^\prime)$ scattering off light nuclei (see Refs.~\cite{Carlson:1998,Bacca:2014,Carlson:2015}
for reviews which include extensive listings of original references).  In the axial sector, however,
this evidence is not as well established, in that discrepancies between experimental data and
theoretical predictions obtained with one-body currents are not as large as in the electromagnetic
case and concern, primarily, the very low momentum and energy transfers of relevance in
$\beta$ decays of very light nuclei~\cite{Schiavilla:1998,Baroni:2018,Pastore:2018}.

Given the above context, the objectives of the present work are twofold: (i) to formulate a quantum Monte 
Carlo (QMC) method for calculating, {\it ab initio}, inclusive muon-capture rates in nuclei in the mass range
$A$=3--12, and (ii) to test our present modeling of the nuclear charge-changing weak current by comparing
theoretical results with available experimental data. This will validate the modeling in a range of momentum
and energy transfers that is intermediate between those relevant, at the low end, in $\beta$ decays and,
at the high end, in neutrino scattering.  We begin by establishing the kinematics
of the process and expressing the rate in a form amenable to a QMC calculation, and then focus
on muon capture in $^4$He as a first practical application of the method.

The muon is captured by the nucleus from
an atomic orbital, and its momentum and energy are denoted by ${\bf k}_\mu$ and $E_\mu$, with
the understanding that $k_\mu \rightarrow 0$, since the muon orbital velocity is of order
$Z\, \alpha \ll 1$ for light atoms.  The muon-neutrino momentum and energy are denoted as
${\bf k}_\nu$ and $E_\nu$ ($E_\nu\,$=$\,k_\nu$), and the masses of the proton and neutron
as $m_p$ and $m_n$.  In the capture process a proton in the initial atom
is converted into a neutron, and energy conservation requires
\begin{equation}
\label{eq:e1}
\Delta m+ E_i= E_\nu + E_f \ , \qquad \Delta m= m_\mu +m_p-m_n \ ,
\end{equation}
where $E_i$ is the internal energy (of electrons and nucleons) of the initial atom, and $E_f$ is
the energy of the final atom including both its internal and recoil energies.  Of course,
binding energies of electrons, at least for light atoms, are of the order of tens of eV's, and therefore negligible when
compared to those of nucleons.

The transition amplitude for capture at leading order is given by
\begin{equation}
T_{fi}=
\frac{G_V}{\sqrt{2}}\, \psi(0)\left[ \overline{u}_{{\bf k}_\nu h_\nu} \, \gamma_\sigma
 \left(1-\gamma_5\right) u_{{\bf k}_\mu s_\mu}\right] j_{fi}^\sigma(-{\bf k}_\nu) \ ,
\label{eq:e6}
\end{equation}
where $G_V$ is the vector coupling constant for semi-leptonic weak interactions
($G_V\,$$\approx$$\,1.1363 \times 10^{-5}$ GeV$^{-2}$~\cite{Hardy:2015}), $u_{{\bf k}_\mu s_\mu}$  and $u_{{\bf k}_\nu h_\nu}$
are the spinors (normalized here as $u^\dagger u\,$=$\,1$) of, respectively, the muon with spin
projection $s_\mu$ and neutrino with helicity $h_\nu$, $j^\sigma_{fi}$ is the matrix element
of the hadronic charge-lowering weak current, 
\begin{equation}
j^\sigma_{fi}(-{\bf k}_\nu)=
\langle-{\bf k}_\nu,f|\int d{\bf x}\, {\rm e}^{-i {\bf k}_\nu \cdot {\bf x} } \,j^\sigma({\bf x}) | i, J_iM_i\rangle \ .
\end{equation}
Since the matrix element $\langle f| j^\sigma({\bf x}) | i\rangle$ is localized over length scales of a few fm's,
the atomic wave function $\psi(x)$ of the muon has been approximated by its value at the origin,
$\psi(0)\,$$=$$\,\left(Z\alpha \mu\right)^3/\pi$ where $\alpha$ is the fine structure constant,
and $\mu$ is the reduced mass of the muon relative to the initial nucleus with $Z$ protons.
Note that the two-component spin-state $\chi_{s_\mu}$ of the muon has been replaced by the spinor, which
is justified in the limit $k_\mu \rightarrow 0$ (and also helpful for carrying out the sums over spins by standard
trace techniques).  Finally, $|i, J_iM_i\rangle$ and $|-{\bf k}_\nu, f\rangle$ are, respectively, the initial nuclear
state with spin and spin-projection $J_iM_i$ and the final nuclear state recoiling with momentum $-{\bf k}_\nu$
with quantum numbers collectively specified by the label $f$.  

The transition rate, when averaged over the spin projections of the initial
nucleus and muon, and summed over those of the final nucleus, is independent
of the $\hat{\bf k}_\nu$ direction and reduces to the well known expression in terms
of Coulomb, longitudinal, electric, and magnetic multipoles (see, for example,
Ref.~\cite{Walecka:2004}).  In the present context, however, we find it convenient to
express this rate (differential in the emitted neutrino energy, but
integrated over the solid angle) in terms of five response functions
\begin{eqnarray}
\frac{d\Gamma}{dE_\nu} &=& \frac{G_V^2}{2\pi}\, |\psi(0)|^2 \, E_\nu^2 \big[ 
R_{00}(E_\nu)+ R_{zz}(E_\nu)  + R_{0z}(E_\nu)\nonumber\\
&&\hspace{1.5cm}  +\, R_{xx}(E_\nu)  - R_{xy}(E_\nu) \big]\ ,
\label{eq:xsw}
\end{eqnarray}
with
\begin{eqnarray}
\label{eq:r1}
\hspace{-0.45cm} R_{00}(E_\nu)&=& \overline{\sum}_{i,f}\delta(\cdots) \, | \, j^0_{fi} (-{\bf k}_\nu)|^2   ,\\
\label{eq:r2}
\hspace{-0.45cm}R_{zz}(E_\nu)&=& \overline{\sum}_{i,f} \delta(\cdots)\,
|\, j_{fi}^\parallel(- {\bf k}_\nu)|^2 ,  \\
\label{eq:r3}
\hspace{-0.45cm}R_{0z}(E_\nu) &=& - \overline{\sum}_{i,f }\delta(\cdots)\, 2 \,{\rm Re}\Big[ 
 \, j^0_{fi}(-{\bf k}_\nu)\,   j_{fi}^{\parallel\,*}(-{\bf k}_\nu) \Big]   ,\\
\label{eq:r4}
\hspace{-0.45cm}R_{xx}(E_\nu)&=& \overline{\sum}_{i,f }\delta(\cdots)\,
|\,{\bf j}_{fi}^\perp(-{\bf k}_\nu )|^2   , \\
\hspace{-0.45cm}R_{xy}(E_\nu)\!&=\!& i \overline{\sum}_{i,f } \delta(\cdots)\, \hat{\bf k}_\nu\cdot \Big[
 \, {\bf j}_{fi}^\perp(-{\bf k}_\nu)  \times {\bf j}_{fi}^{\perp\,*}(-{\bf k}_\nu)\Big] ,
\label{eq:r5}
\end{eqnarray}
where we have introduced the unit vector $\hat{\bf k}_\nu\,$=$\, {\bf k}_\nu/E_\nu$, the longitudinal
and transverse components of the current, respectively $j_{fi}^\parallel\,$=$\hat{\bf k}_\nu\cdot{\bf j}_{fi}$
and ${\bf j}_{fi}^\perp\,$=${\bf j}_{fi}-\hat{\bf k}_\nu \,j_{fi}^\parallel $, and the short-hand notation
$\delta(\cdots)$ for the energy-conserving $\delta$-function resulting from Eq.~(\ref{eq:e1}).  The
bar over the summation symbol implies the average over (nuclear) spin projections indicated
earlier.

As they stand, a calculation of these response functions by QMC
methods~\cite{Carlson:2015,Lovato:2013,Lovato:2015,Lovato:2016,Lovato:2018}
is not possible, since the lepton momentum and energy transfers, respectively ${\bf q}$ and $\omega$,
in the weak capture (like in a photo-absorption process) are not independent variables; indeed,
${\bf q}=- E_\nu \, \hat{\bf k}_\nu$ and $\omega=m_\mu -E_\nu$.  To circumvent this difficulty,
we consider instead (in a schematic notation)
\begin{equation}
R_{\alpha\beta}(q,\overline{\omega})=\overline{\sum}_{if} 
\delta(\overline{\omega}+E_i-E_f)\, O^\alpha_{fi}({\bf q}) \, O^{\beta *}_{fi}({\bf q}) \ ,
\end{equation}
with ${\bf q}\,$=$\,-E_\nu \, \hat{\bf k}_\nu$ and $\overline{\omega}$ taken as independent variables.
We carry out the Laplace transform 
\begin{eqnarray}
E_{\alpha\beta}(q,\tau)&=&\int_0^\infty d\,\overline{\omega}\, {\rm e}^{-\tau \,\overline{\omega}}\,
R_{\alpha\beta}(q,\overline{\omega}) \nonumber\\
&=&\overline{\sum}_i \langle i |O^{\beta\dagger} ({\bf q}) {\rm e}^{-\tau\left(H-E_i\right)}  O^\alpha({\bf q})|i\rangle \ ,
\label{eq:e11}
\end{eqnarray}
by evaluating the expectation value in the second line above with stochastic techniques~\cite{Carlson:1992},
invert the resulting Euclidean response function $E_{\alpha\beta}(q,\tau)$ by maximum-entropy methods~\cite{Lovato:2015}
to obtain back $R_{\alpha\beta}(q,\overline{\omega})$, and finally interpolate the latter at
$\overline{\omega}\,$=$\,\omega+m_p-m_n\,$=$\,\Delta m-E_\nu$ to determine the response
$R_{\alpha\beta}(E_\nu)$ of interest here.  No approximations are made beyond those inherent to
the modeling of the nuclear Hamiltonian and weak current; in particular interaction effects in the
discrete and continuum spectrum of the final nuclear system are fully and exactly accounted for.

The dynamical framework adopted in the present work is based on a realistic Hamiltonian
including the Argonne $v_{18}$ two-nucleon~\cite{Wiringa:1995} (AV18) and Illinois-7
three-nucleon~\cite{Pieper:2008} (IL7) interactions, and on realistic charge-changing weak
currents with one- and two-body terms, see Ref.~\cite{Shen:2012} for a recent overview
and a listing of explicit expressions.  The (vector and axial) one-body terms $j^\sigma_{1{\rm b}}$
follow from a non-relativistic expansion of the single-nucleon (charge-changing) weak current,
in which corrections proportional up to the inverse-square of the nucleon mass are retained.
The two-body currents $j^\sigma_{2{\rm b}}$ consist of contributions associated with (effective)
$\pi$- and $\rho$-meson exchanges, and $N$-to-$\Delta$ excitation terms, treated in the static
limit.  In the axial component, a $\rho\pi$ transition mechanism is also included.  Configuration-space
representations of these currents (used in the actual calculations below) are regularized by a
prescription which, albeit model dependent, is nevertheless designed to make, by construction,
their short-range behavior consistent with that of the two-nucleon interaction---the
AV18.  In the $N$-to-$\Delta$ axial current, the value for the transition (axial) coupling
constant is determined by reproducing the measured Gamow-Teller matrix element contributing to
tritium $\beta$-decay~\cite{Shen:2012} (within the present dynamical framework).  The level of
quantitative success these currents have achieved, when used in combination with the
AV18+IL7 Hamiltonian, in accurately predicting many electroweak properties of s- and p-shell nuclei up to
$^{12}$C is illustrated in Refs.~\cite{Carlson:1998,Carlson:2015} and references therein.
\begin{widetext}
\vspace{-0.5cm}
\begin{center}
\begin{table}[bth]
\begin{tabular}{ c|c|c||c|c||c|c||c|c||c|c|c||c|c}
                   &  V-1b & V-2b & A-1b & A-2b & CC-1b  & CC-2b & $\widetilde{\rm CC}$-1b & $\widetilde{\rm CC}$-2b &Exp~\cite{Bizzarri:1964}& Exp~\cite{Auerbach:1967} & Exp~\cite{Block:1968} & Th~\cite{Caine:1963} & Th~\cite{Walecka:1975} \\
                   \hline
$\Gamma$(s$^{-1})$ &  $ 65 \pm 1 $  & $ 73 \pm 1 $ & $ 171 \pm 6$ & $ 200 \pm 6$ & $ 265\pm 9$ & $ 306\pm 9$ & $310\pm 12$ & $355\pm 12$ & $336\pm 75$  &$ 375^{+30}_{-300}$ & $364\pm46$  & $345\pm 110$ & 278  \\
\hline
\end{tabular}
\caption{The inclusive muon rates in $^4$He obtained by including one-body (1b) only
and both one- and two-body (2b) terms in the vector (V) and axial (A) components
of the charge-changing (CC) weak current.  The 1b and 2b rates obtained with the full CC current
and the $\widetilde{\rm CC}$ current without the induced pseudoscalar term
are compared to available experimental values and older theoretical estimates.}
\label{tb:tb1}
\end{table}
\end{center}
\vspace{-1cm}
\end{widetext}

Having set-up the formalism and specified the dynamical framework, we now proceed to discuss
an application of the method to muon capture in $^4$He.  As noted by Measday in his review~\cite{Measday:2001},
the only available measurements of the total rate are from experiments in the 1960's with helium bubble
chambers and helium gas scintillating targets~\cite{Bizzarri:1964,Auerbach:1967,Block:1968}, and have
large errors, see Table~\ref{tb:tb1}.  The only theoretical estimates we are aware of are from Caine and
Jones~\cite{Caine:1963} and Walecka~\cite{Walecka:1975}; the former based on closure approximations
is rather uncertain, while the latter obtained with the Foldy-Walecka sum rules for the giant dipole excitation
turns out to be remarkably close to the value we calculate almost 50 years later!

The calculation of the $^4$He Euclidean responses in Eq.~(\ref{eq:e11})
is carried out with Green's function Monte Carlo (GFMC)
methods~\cite{Lovato:2013,Lovato:2015,Lovato:2016,Lovato:2018}
similar to those used in projecting out the exact ground state of a Hamiltonian from a trial
state~\cite{Carlson:1987}.  It proceeds in two steps.  First, an unconstrained imaginary-time
propagation of the initial bound state state $| i\rangle$, represented here by an accurate
variational Monte Carlo (VMC) wave function (rather than its exact GFMC counterpart),
is performed and saved.  Next, the states $O^\alpha({\bf q})|i\rangle$ are evolved in imaginary
time following the path previously saved.  During this latter imaginary-time evolution, scalar
products of ${\rm exp}\left[-\left(H-E_i\right) \tau_i\right]O^\alpha({\bf q}) |i\rangle$
with $O^\beta({\bf q}|i\rangle$ are evaluated on a grid of $\tau_i$ values, and
from these scalar products estimates for $E_{\alpha\beta}(q,\tau_i)$ are
obtained.  The statistical errors associated with the GFMC evolution
remain modest, even at values of $\tau$ as large as $0.1$ MeV$^{-1}$,
the endpoint of the $\tau$-grid.  Maximum entropy methods are employed
``to invert'' $E_{\alpha\beta}(q,\tau)$ and obtain the corresponding
$R_{\alpha\beta}(q,\overline{\omega})$~\cite{Lovato:2015}.
Their implementation is briefly summarized in the supplemental material.

\begin{figure}[bth]
\includegraphics[width=\columnwidth]{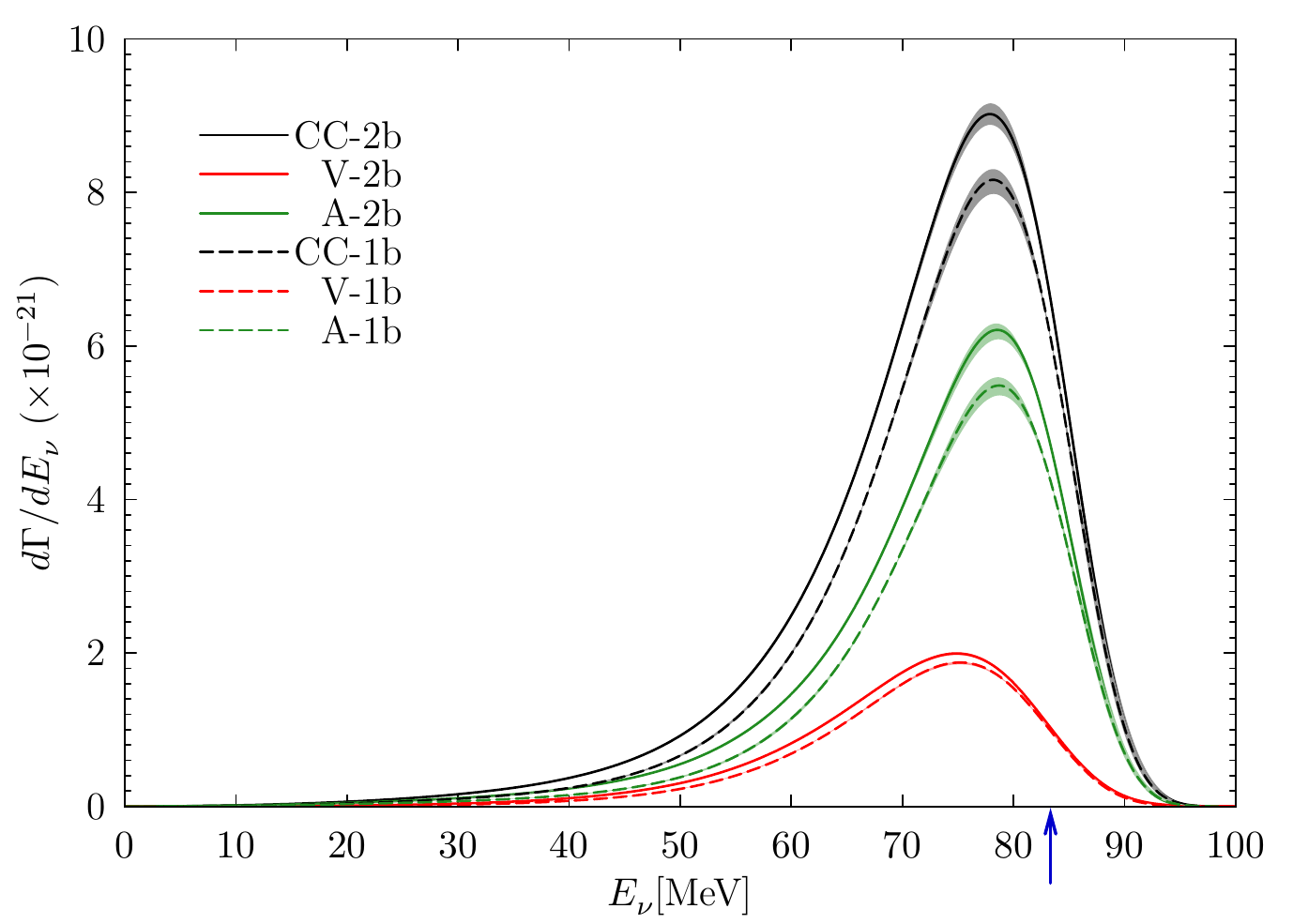}
\caption{(Color online). The differential rates obtained with
one-body (1b) only and both one- and two-body (2b) terms in the vector (V) and axial (A) components
of the charge-changing (CC) weak current, and full CC current, are displayed as function of the
$\nu_\mu$-energy in the allowed kinematical range.  The theoretical
uncertainites resulting from combining statistical errors in the GFMC calculation with errors associated with
the maximum-entropy inversion of the imaginary-time data are shown by the bands.
The arrow indicates the kinematically maximum allowed $E_\nu$, see text for further explanations.}
\label{fig:f1}
\vspace{-0.25cm}
\end{figure}
Predictions for the total rate in $^4$He are compared to the experimental values and older theoretical
estimates mentioned above in Table~\ref{tb:tb1}, and the differential rates as functions of the energy of the
muon neutrino emitted in the capture are shown in Fig.~\ref{fig:f1}.  Results obtained by considering
only the vector (V) or axial (A) components of the charge-changing (CC) weak current and by
including one-body (1b) terms only or both one- and two-body (2b) terms in these currents are listed
in Table~\ref{tb:tb1}, and displayed in Fig~\ref{fig:f1}, separately.  Note that the response function
$R_{xy}(E_\nu)$ in Eq.~(\ref{eq:r5}) involves interference between the matrix elements of the V and A
currents, and therefore only contributes when both are present.  As a consequence, $\Gamma({\rm CC})\!$
$\ne$$\, \Gamma({\rm V})+\Gamma({\rm A})$; indeed, this V-A interference leads to an increase
in the $\Gamma({\rm V})+\Gamma({\rm A})$ result by $\approx 10$\% in both the 1b- and 2b-based
calculations.

In the $^4$He capture, the neutrino energy is in the range
$0\le E_\nu \le E_\nu^{\rm max}\,$$\approx$$\,83.6$ MeV; however, the distribution,
on account of the $E^2_\nu$-weighing factor present in the expression for $d\Gamma/dE_\nu$,
is skewed towards the high end, confirming the expectation that the energy release in
the capture process is converted primarily into energy for the emitted neutrino~\cite{Measday:2001}
with the remaining balance being absorbed by the final nuclear system.  In the present
case, since $^4$H is not bound, the possible final breakup channels are $^3$H+$n$ (3+1),
$^2$H+$2\, n$ (2+2), and $^1$H+$3\,n$ (1+3), which have slightly different thresholds.  While the
contributions of these channels are fully accounted for here, they
cannot be individually identified over the allowed $E_\nu$ range---a limitation
intrinsic to the present method and apparent from Eq.~(\ref{eq:e11}), which
relies on closure to remove the sum over final states.  Despite relying on the closure approximation, Caine
and Jones~\cite{Caine:1963} estimated the branching ratios into the 3+1, 2+2,
1+3 channels to be, respectively, 97.75\%, 2\% and 0.25\%.

A related issue has to do with the behavior of the response functions in the threshold region
$E_\nu \lesssim E_\nu^{\rm max}$.  The kinematical constraint that $R_{\alpha\beta}(E_\nu)$
vanish for $E_\nu$ larger than $E_\nu^{\rm max}$ is not imposed when performing the
inversion (see supplemental material).  Even though relatively high values of $\tau \le \tau^{\rm max}\,$=$\,0.1$ MeV$^{-1}$
are calculated by GFMC, the maximum-entropy procedure we utilize still produces
some strength beyond $E_\nu^{\rm max}$, as is apparent from Fig.~\ref{fig:f1}.
However, the integrals of $d\Gamma/dE_\nu$, when evaluated over the whole $E_\nu$-range
including the unphysical region, remain stable to within 1\% for $\tau^{\rm max}\,$=$(0.1,0.08,0.05)$
MeV$^{-1}$.

In Table~\ref{tb:tb1} we also list the results for the 1b and 2b total inclusive rates (indicated
as $\widetilde{\rm CC}$) obtained with an incomplete CC weak current in which the
term proportional to the induced pseudoscalar form factor $G_{PS}(q^2)$ (in the axial sector) is
ignored.  The effect is significant: retaining this term reduces the $\widetilde{\rm CC}$ values by
$\approx\,$15\% (14\%) in the 1b (2b) calculations.  The parametrization for $G_{PS}(q^2)$
adopted here~\cite{Shen:2012} is consistent with the recent determination of this form factor by the
MuCap collaboration~\cite{Andreev:2013}. It also leads, in an accurate {\it ab initio} calculation based on
essentially the same dynamical inputs adopted here~\cite{Marcucci:2011}, to a prediction for
the $^3$He($\mu^-,\nu_\mu$)$^3$H total rate that is agreement with the (remarkably precise)
measurement of Ref.~\cite{Ackerbauer:1998}, 1496(4) s$^{-1}$.  Thus, muon capture provides
a sensitive test of the $G_{PS}(q^2)$ form factor at low momentum transfers.  By contrast,
this observable is only very marginally affected (at a fraction of a 1\% level) by changes in the
parametrization of the nucleon axial form factor, as we have explicitly verified by calculating how
the total rate changes when the cutoff $\Lambda_A$ is varied by $\pm 10$\% about its central
value of $\Lambda_A\,$$\approx$$\, 1$ GeV.  The reason is that $G_A(q^2)\,$=$\,g_A\left[ 1+2\,
q^2/\Lambda_A^2+ \cdots\right]$, and $q^2/\Lambda_A^2\ll1$ in the allowed kinematical region.

In this letter, we have formulated an {\it ab initio} QMC method for calculating inclusive muon-capture
rates on light nuclei (mass number $A\le 12$), and have presented, as a first application, a calculation
of the total rate in $^4$He.  The predicted value is consistent with the lower range of available
experimental determinations (see Table~\ref{tb:tb1}).  However, these measurements from bubble
chamber experiments of the late 60's have large errors, making it impossible to establish, at a
quantitative level, the validity of the model for the nuclear charge-changing weak current we have
adopted here.  We hope the present work will motivate our colleagues to carry out
a new experiment on $^4$He.

Future plans in this area include (i) the application of the method to other (light) nuclei, especially
in cases where more accurate data are known~\cite{Measday:2001}, and (ii) its extension to
more fundamental dynamical approaches based on interactions and electroweak currents
derived from chiral effective field theory.  The presence of discrete states in the final nuclear
system substantially complicates the calculation of the capture rate, since the imaginary-time
response functions $E_{\alpha\beta}(q,\tau)$ would have to be evaluated at large enough values
of $\tau$ to reliably resolve the contributions of these states in the threshold region of the
corresponding ``inverted'' $R_{\alpha\beta}(E_\nu)$---that is, when $E_\nu$ is close to the
maximum kinematically allowed value.  A similar issue arose in the calculation of
the longitudinal and transverse electromagnetic response functions of $^{12}$C~\cite{Lovato:2016}.

We acknowledge the support of the U.S.~Department of Energy, Office of Science, Office
of Nuclear Physics, under contracts DE-AC02-06CH11357 (A.L.~and N.R.) and DE-AC05-06OR23177 (R.S.),
as well as the support by the NUclear Computational Low-Energy Initiative (NUCLEI) SciDAC project
(N.R.) and by the Fermi Research Alliance, LLC, under contract DE-AC02-07CH11359 with
the U.S.~Department of Energy, Office of Science, Office of High Energy Physics (N.R.).
Under an award of computer time provided by the INCITE program, this research used resources of the Argonne
Leadership Computing Facility at Argonne National Laboratory, which is supported by the Office of Science of the
U.S. Department of Energy under contract DE-AC02-06CH11357.
Computational resources provided by the National Energy Research Scientific Computing Center (NERSC)
are also gratefully acknowledged.
%
%

%
%
%
%
\end{document}